\documentclass{eas_vlti}
\usepackage{graphicx}
\TitreGlobal{What can the highest angular resolution bring to stellar astrophysics?}
\begin{document}

\title{Star Formation at milli-arcsecond resolution} 
\runningtitle{Star Formation at milli-arcsecond resolution} 
\author{Ren\'e D. Oudmaijer}\address{School of Physics \& Astronomy,
  University of Leeds, Woodhouse Lane, Leeds LS2 9JT, UK}
\author{Willem-Jan de Wit}\address{European Southern Observatory, Casilla 19001, Santiago 19, Chile
}
%
%
\begin{abstract}

This chapter discusses the use and possibilities of optical and
infrared interferometry to study star formation. The chapter starts
with a brief overview of the star formation process and highlights the
open questions from an observational point of view. These are found at
the smallest scales, as this is, inevitably, where all the action such
as accretion and outflows, occurs. We then use basic astrophysical
concepts to assess which scales and conditions can be probed with
existing interferometric set-ups for which we use the ESO/VLTI
instrument suite as example. We will concentrate on the more massive
stars observed at high resolution with continuum interferometry.
Throughout, some of the most recent interferometric results are used
as examples of the various processes discussed.

\end{abstract}
\maketitle
\section{Introduction}

The basic principle of the gravitational collapse of a molecular cloud
leading to the formation of a young star has been known since the
ground-breaking work by Jeans in the early 20$^{\rm th}$
century. Indeed, Jeans' major insight has led to significant progress
in our understanding of the star formation process. The prevailing
formation scenario for low mass stars, was outlined in detail in their
standard work by \cite{shu_1994}. In short, after the dense, cold
molecular cloud collapses, conservation of angular momentum implies
that the system rapidly spins up, leading to the formation of a
circumstellar disk around the central star. Matter will continue to
accrete onto the star through this accretion disk, while bi-polar jets
emanate from the polar regions of the newly formed star.  The disk not
only feeds the star, but also acts as birth place of possible
planets. This paradigm of collapsing clouds forming stars has been
observationally verified and forms the basis of our understanding of
low mass star formation.

For more massive stars, progress has been slower, both theoretically and
observationally. A key difference is that the lower mass, Sun-like, stars
sustain magnetic fields, and grow via magnetically controlled accretion
through a disk. In contrast, higher mass stars have radiative envelopes and
are not magnetic. They are thus not necessarily expected to form in a similar
manner to low mass stars.  In addition, the larger masses and therefore larger
stellar luminosities involved pose their own difficulties as the stars'
enormous power output results in a strong radiation pressure which may halt
the accreting material altogether and prevent the stars to grow to very high
masses.  Recently \cite{krumholz_2009} were able to show from high resolution
computations that massive stars can indeed form by disk accretion. The
radiation driven outflows are channeled through the bipolar cavity and
consequently the disk undergoes less radiation pressure.
Confirmation of this formation scenario revolves around the properties of
the circumstellar material close to the star. Inevitably, this requires
observations at the highest spatial resolution.

Observationally, the study of massive young stars is hampered by the
fact that they are much rarer than lower mass stars, and therefore on
average further away, requiring high resolution studies to probe the
accretion region. This is one of the reasons that direct imaging of
these objects was not possible for a long time.  As an alternative,
astronomers resorted to the next best diagnostic for studying the
circumstellar material instead, the Spectral Energy Distribution
(SED).  The trouble however is that model fits of SEDs are not unique. They
can be readily explained by both disks or spherical envelopes, and as
a result very different star formation scenarios can be invoked based
on the same data.  The ambiguity associated with indirect methods is
well illustrated by the lively debate in the nineties about the nature
of the circumstellar material around the intermediate mass pre-Main
Sequence Herbig Ae/Be stars. The debate concerned whether Herbig Ae/Be
stars were surrounded by disks or not (see e.g. \cite{waters_1998} for
an overview at that time).  Headway was only made when
\cite{mannings_1997} detected rotating disk-like structures around
Herbig Ae/Be stars, and the presence of disks was seemingly
established. Not much later, \cite{miro_1999} showed that the
simultaneous presence of both disk and envelope can also explain the
SED; the inner, warm, disk contributes to the near-infrared (NIR)
excess, while a cooler spherical envelope dominates the longer
wavelength excess emission. The case of disk accretion for Herbig
Ae/Be was not yet settled however, as the large (200-600 au) disk-like
structures observed at mm wavelengths are too far from the star to
probe accretion.

It is important to note here that on several occasions flattened structures
have been observed towards massive young stars, and these were often
identified with disk accretion scenarios. However, this is a misconception,
many such disk tracers only probed regions far out and thus can not at all
be used to study the accretion process.  This issue is often overlooked.

Not surprisingly, progress often goes hand in hand with technological
advances, and star formation is not an exception. In particular the advent of
milli-arcsecond resolution observations offers the prospects to study the
inner regions where the accretion occurs.

\subsection{Properties of the circumstellar material}

Before we start discussing observations, and their interpretation,
that have been performed at the highest resolution, it may be
appropriate to first discuss a model of the geometry that we expect
around these stars. In order to do so, we will first introduce some
terminology and explain the various types of star in the framework of
the classical Lada classification of the various stages of the
formation of a star (Lada, 1987). In short, the class I sources are
very cool: not much, if any, heating of the molecular cloud occurs and
these objects are only visible at infrared and longer wavelengths. At
a later stage, the class II sources show a flat spectrum, light from
the accretion disk and the envelope dominate the SED, at near-infrared
wavelengths and longer. The class III sources have dispersed these
envelopes in their entirety and the SED is dominated by the star and
disk.  The Class II and III sources, being bright at near- and
mid-infrared wavelengths ($K$ - $N$ band), have their peak energy
output in the ESO/VLTI wavelength range, while the Class I sources can
be studied at the $L, M$ and $N$ bands. Hence the ESO/VLTI operating
wavebands are particularly well matched to the young stars (or vice
versa).

The types of low mass sources (M $<$ 2-3 M$_{\odot}$) that we can
study are the pre-main sequence T Tauri objects which are type II/III
sources. Their counterparts at intermediate mass (2-3 M$_{\odot}$ $<$
M $<$ 10-15 M$_{\odot}$) are the pre-main sequence Herbig Ae/Be
stars. At higher masses there is no obvious, observable, pre-Main
Sequence phase, as the stars evolve much faster and arrive on the Main
Sequence while still embedded in their envelopes. In a way, the
objects under consideration, the Massive Young Stellar Objects (MYSO)
can be compared to the class I objects.

 A picture of their circumstellar environment is provided in Fig.~1. The star
 is embedded in a dense envelope, surrounded by an accretion disk, while the
 radiation pressure from the star drives a flow which carves a bipolar cavity
 in the envelope.

\begin{figure}
\begin{center}
  \includegraphics[width=0.6\textwidth]{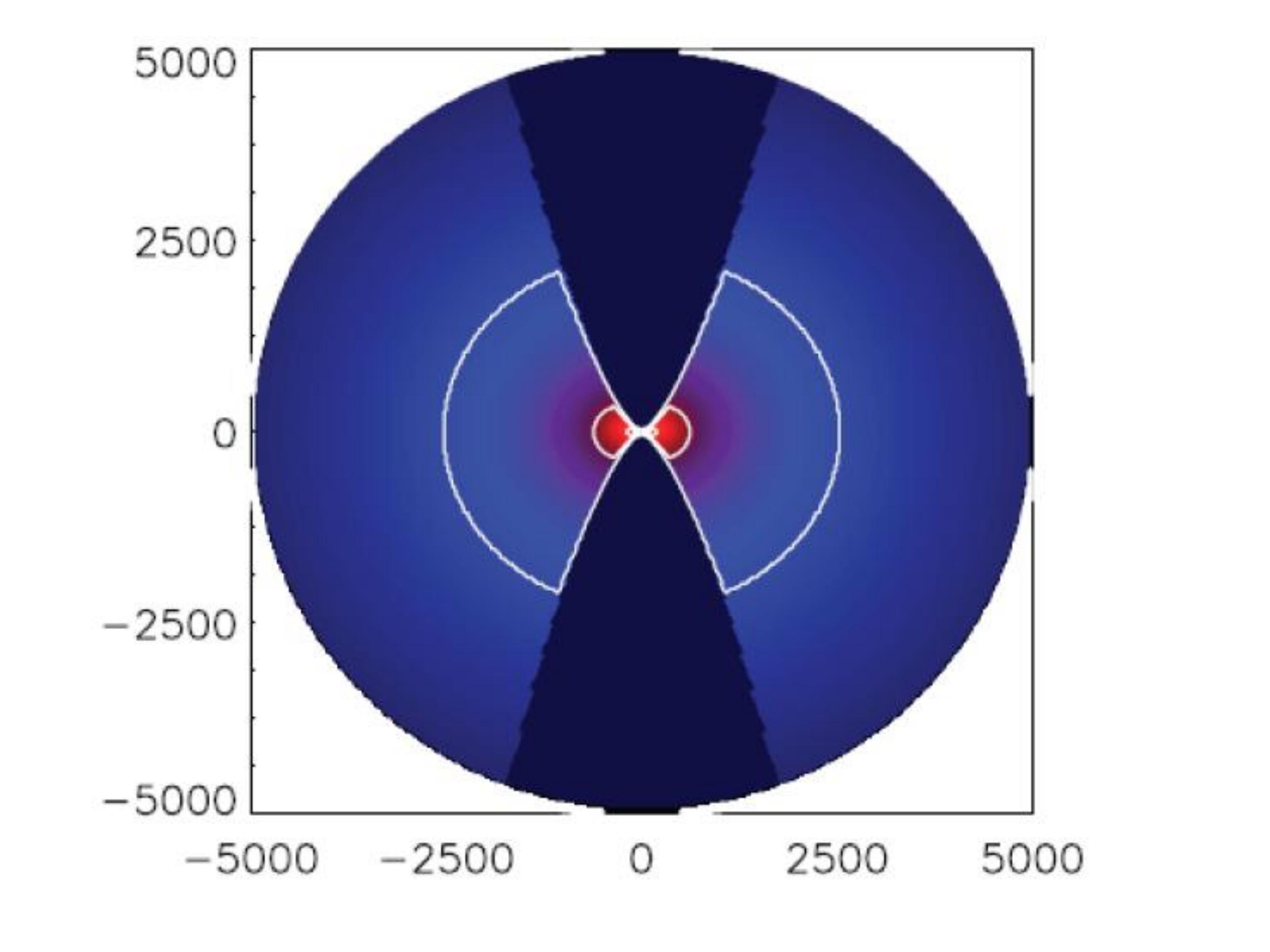}
  \caption{A model representation of the circumstellar matter of a
    Massive Young Stellar Object. The central star is embedded in a
    dense dusty environment. The bi-polar flow has carved a cavity,
    while the equatorial condensation may signpost a dense accretion
    disk. The scales are in au. From Whitney et al. (2003)}
\end{center}
\end{figure}

\subsection{Observations of the circumstellar material}

Many breakthroughs in this field over the last years have been made
with the new generation optical/infrared interferometric
instrumentation. These made observations of the inner disks and the
bases of the outflows at milli-arcsecond scales possible. This is at
scales which are orders of magnitude smaller than the outer disks
detected and probed by infrared and (sub-)mm observations obtained
thus far.  The inner disks were first observed using single baselines
which allowed basic size measurements. It was found that the inner
boundaries of the disks get larger as a function of luminosity,
consistent with the notion that the dust sublimation radius is
measured (as documented by e.g. \cite{millangabet_2007}, and discussed
later). However, not all Herbig Ae/Be stars follow this relationship,
and some objects turned out to be smaller than predicted. This has led
to several suggestions for the nature of the near-infrared
emission. Competing ideas include that the near-infrared emission
comes from optically thick gas (\cite{kraus_2008}), or that it is due
to refractory grains which can survive higher temperatures
(\cite{benisty_2010}).  Full model-independent images of several
Herbig Ae/Be stars have now been published, proving beyond doubt that
the objects are surrounded by disks (e.g. \cite{benisty_2011} and
references therein).  In parallel, the next step is to fully
understand and parametrize the properties of the disks in order to
follow the formation and evolution of intermediate and massive stars.
An early example has been published by \cite{weigelt_2011} who took
advantage of the high spectral resolving power of VLTI/AMBER ($R \sim
12000$) and both spectrally and spatially resolved the Br$\gamma$ line
of the Herbig Be star MWC 297 which has a spectral type of B1.5Ve. By
fitting the spectral line, they suggest that the emission is due to a
disk-wind.

Finally, there are the more embedded Massive Young Stellar Objects
with masses exceeding 10-15 M$_{\odot}$, these are on average even
further away than Herbig Ae/Be stars, with as added complication that
they are optically invisible due to the large extinction.  Using AMBER
data at 2$\mu$m, \cite{kraus_2010} present tantalizing evidence for
disks around these most massive young stars as well.

In this chapter we will first discuss what we can probe at the
smallest scales for young stars. We then discuss the power and
limitations of optical/infrared (OIR) interferometric data in
understanding the properties of these objects. We continue describing
the efforts in interpreting the data supported by radiative transfer
models and highlight the various pitfalls. We conclude the chapter
with a number of exercises.  We will not be able to discuss every
aspect of star formation, and will focus here on the more massive
stars and high resolution OIR continuum interferometry. A very recent
overview of the field can be found in the review papers of the
Protostars and Planet VI conference held in Heidelberg in 2013
(\underline{www.ppvi.org}). This chapter also touches upon many topics
discussed in this book, of which the contributions by Bonneau et al.
(binary stars), and Meilland \& Stee (disk diagnostics) are of note.

\section{What we  are probing at small scales}

At the wavelengths the VLTI operates (AMBER: $H, K$, MATISSE: $L, M$,
MIDI: $N$, hence 1.6 $\mu$m to 10 $\mu$m), the observed flux from
young stars is dominated by thermal emission from dusty particles,
whose temperature determines the peak wavelength. The dust temperature
is a function of distance to the star which we can relate to angular
size scales using some basic astrophysical principles.

Assuming that a dust grain acts like a black body, and re-radiates all
the energy it absorbs, it can be shown that the temperature of a dust
grain depends only on its distance from the star and the star's
luminosity.  Using Wien's displacement law ($\lambda T \approx 3 \times
10^{-3}$ mK), we can determine at which wavelengths the dust
radiation dominates, and therefore relate astrophysical size scales to
the observational set-ups required. We leave the specifics of the
derivation of the relationships between the dust temperature, $T_d$,
the stellar parameters (temperature and radius, $T_{*},R_{*}$), and
the distance of the dust to the star, $d$ as an exercise to the reader
at the end of this chapter. The result can be written as:

\begin{equation}
 T_d = \sqrt{\frac{R_*}{2d}} T_* 
\label{dusty}
 \end{equation}

Intuitively, this equation makes sense, the higher the star's
temperature and radius, the more energy it emits and hence more energy
is absorbed by a dust particle.  In contrast, the larger the distance
between star and dust, less energy is absorbed.  Given that $L_*
\propto R_*^2T_*^4$, we can see that $ T_d^2 \propto \sqrt{L_*}/{d}$.
The radius of the dust grains cancel out in the derivation. This
is because the energy absorbed by the grain and that radiated away
by it are both  dependent on the dust grain radius squared.

Knowing the dust temperature, we can then obtain an estimate of the
distances probed at the various emitted wavelengths. An interesting
early result demonstrating this relationship was obtained by
Millan-Gabet et al. (2007). Using single-baseline optical
interferometry, they measured the angular sizes of the dust emitting
regions of a sample of young stellar objects at the $K$ band. This
wavelength probes, to first order, the inner parts of the disks at the
dust sublimation temperature, the highest possible temperature below
which dust can exist.  The relationship above then reduces to $d
\propto \sqrt{L_*} $, which is precisely what these authors observed
for a sample of objects. Remarkably, the luminosities of their objects
spanned five orders of magnitude. Recently, it has been found that
Active Galactic Nuclei follow this relationship too
(\cite{Kishimoto_2011}), extending the observed trend to galactic
scales.

Let us now get a feel for the physical scales involved by evaluating
Eq.~\ref{dusty} for a real Massive Young Stellar Object. Assuming the
central star can be approximated by a Main Sequence star, a typical
MYSO has $R_* = 8.4 R_{\odot}$ and temperature 35000 K (e.g. W33A in
\cite{dewit_2007}). For a typically high dust temperature of 1000 K,
we get

$$ d = \left(\frac{T_*}{T_d}\right)^2 \left( \frac{R_*}{2} \right) = 24 {\rm
    \, \, \, au} $$

\begin{figure}
\begin{center}
  \includegraphics[width=0.6\textwidth]{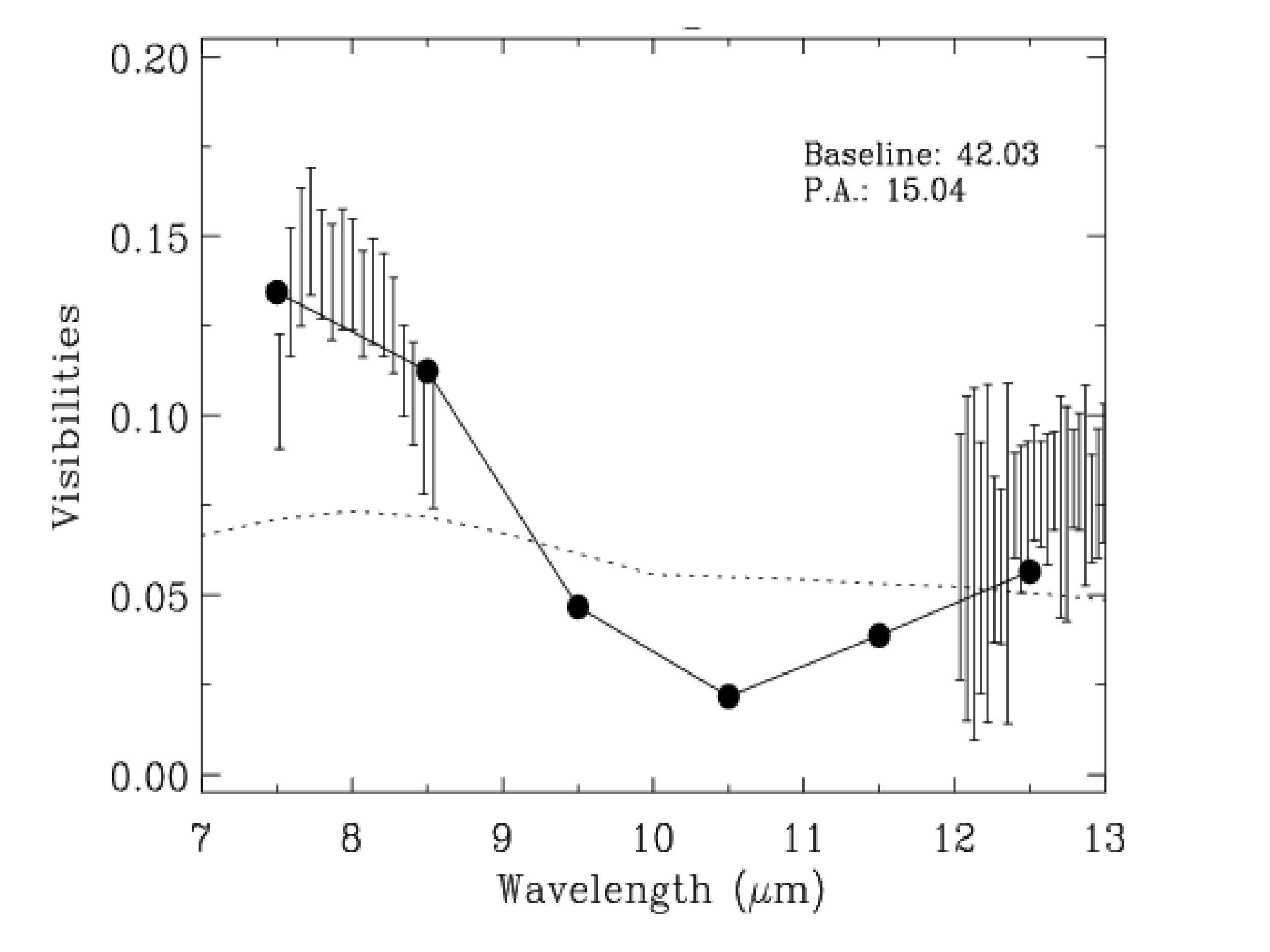}
  \caption{The visibility of the Massive Young Stellar Object W33A
    from 8-13$\mu$m as seen by MIDI on one baseline. The visibilities
    decrease with increasing wavelength, while the central silicate
    absorption at 10 $\mu$m is so deep, that no flux nor visibilities
    could be reliably measured. The lines denote various model fits to
    the data. Figure taken from de Wit et al. (2010).}
\end{center}
\end{figure}

At a distance of 3.8 kpc, this corresponds to 6 milli-arcseconds,
angular scales for which VLTI/AMBER is excellently suited and the
success of the application of the VLTI, and other arrays, to Herbig
Ae/Be stars is testament to this.  However, with the exception of
preciously few published stars, many of these objects are too faint to
be observed at the shorter wavelengths. The situation is even worse
for the more embedded MYSOs. They are much brighter at longer
wavelengths and can often only be observed at the MIDI $N$ band.

However, observing at longer wavelengths introduces a complication,
because the spatial resolution becomes worse at longer wavelengths as
the resolution, $R$ is proportional to
$\frac{\lambda}{2B}$. Therefore, one could argue that MYSOs would
become progressively less resolved at these longer
wavelengths. However an interesting property of these objects is that
they become larger with increasing wavelength, as these longer
wavelengths trace cooler material.  Using Wien's law, we find the MIDI
wavelength range is sensitive to warm dust at 300 K.  As shown above
when we considered the dust which absorbs radiation and re-emits it as
a black body, the distance increases with the inverse of the
temperature squared, and this dust is located $\sim$ 60 mas from the
star.  So, the saving grace in this situation is that the object is
actually more resolved for the same set-up and baseline $B$ at longer
wavelengths. The distance from the star increases with $\lambda^2$ (as
$\lambda T$ is constant), and the ``resolvedness'' (for want of a
better term) of these objects increases with $\lambda$.  An
illustration of this effect is provided in Figure 2. The figure shows
the visibilities of a MIDI spectrum of a MYSO (W33A, data taken from
\cite{dewit_2010}). The spectrum covers the wavelength range from 8 to
13 $\mu$m. The 10$\mu$m silicate absorption feature is so strong that
no correlated flux could be measured in this line, but it is clear
that the visibilities decrease as a function of wavelength, as
predicted.

\section{Interpreting the interferometric data}

The discussion above illustrates both the power of high spatial
resolution interferometry, but also its limitations. The single
baseline visibilities allow for the determination of size scales and
multiple baselines, or even model independent images, enable us to
determine the geometry of the dusty material. However, astrophysical
properties can not be derived from  geometric information alone.
For example, the analysis of the largest published MIDI dataset of
MYSOs by Boley et al. (2013) is necessarily rudimentary and the
authors can not conclude from their data and geometric analysis which
particular structure is responsible for the $N$-band emission. Indeed,
in-depth studies of individual objects using computer (and labour)
intensive interpretation are needed to make progress. It is the
combination of observations of large samples underpinned by in depth
studies that allow progress to be made.

\subsection{Modelling the Spectral Energy Distribution}

One of the main aims in YSO research is to determine the parameters of
the circumstellar material, the dust mass, its distribution and
geometry.

A very powerful instrument in the astronomer's toolbox is the
modelling of the spectral energy distribution (SED) of an object. The
method is fully based on the principles outlined earlier. The distance
of a dust grain to the star and the star's properties determine the
temperature of the dusty particle which emits as a black body. The
larger the particles, more flux will be absorbed and re-radiated, so
the particle size, or even size distribution, determines the total
outgoing infrared flux. The emergent radiation from a source embedded
in a dust cloud is then the sum of the radiation of all dust particles
that we can see. The challenge is to find which combination of star
and dust particles can reproduce the SED.

Sometimes the infrared emission has the same shape as the Planck
function, implying that effectively the dominant dust emission can be
reproduced by a thin shell where all dust particles have the same
temperature. If the infrared spectrum is flatter (or steeper), a
different grain density as a function of distance from the star can be
invoked. In the case of, for example, constant mass loss or accretion,
the density follows a powerlaw $\rho(r) \propto r^{-2}$.

There are many models available that allow one to compute the SED from
a star embedded in a dusty environment (e.g. \cite{mrr},
\cite{whitney_2003}; see also \cite{ivezic_1997}). The basics of these
models are similar, and the main differences are the levels of detail
such as those of the stellar parameters (e.g. black body vs. stellar
model), dust parameters (e.g. size distribution vs. single size, dust
composition which affects the absorption and emission properties) and
the implementation of the geometry (see later). Naturally, each
refinement comes at the price of more input parameters. Some of these
can be reliably set based on other information if available. These
include for example the spectral type of the star when a stellar model
is used as input, or the dust composition from an infrared spectrum,
while others are free parameters that can only be determined through
fitting. Depending on the complexity of the underlying model, many
models can be computed fairly quickly on standard laptops and
desktops.
 
In order to determine the input parameters of a dust model (and
ultimately the properties of the dusty envelope) that provides the
best fit to the SED one can use a formal fitting procedure, where the
minimum $\chi^2$ value is found by searching the entire parameter
space. When the number of free parameters is large, an inordinate
amount of computing time would be required to sample all possible
combinations. In practice, minimum search algorithms such as {\sc
  amoeba} are employed to reduce the computations required by
specifically zooming in on the regions where the minimum $\chi^2$ is
located. An alternative is to compute a huge grid of models, and
choose the best fitting one to the observed SED. An advantage is that
this necessitates only one (long) run and multiple SEDs can be fitted
in a short amount of time, whereas a disadvantage is that the grid does not
necessarily cover all possible combinations of parameters.  A good
example is provided by \cite{Robitaille} who published a large grid of
model SEDs, and posted the models and a bespoke fitting routine on the
internet.

With the caveat that the number of free fit parameters can be large,
it is in practice almost always possible to derive the number of
particles, their density distribution and thus mass of the dusty
envelope, from fitting the SED alone. 

So far, we have not discussed the geometry of the material in terms of
modelling the SED. Indeed, in many cases, spherical symmetry is
assumed from the outset (e.g. in the case of evolved stars). As
explained in the introduction, the SED modelling by itself can not
discriminate between various spatial distributions of the dust.  The
infrared emission is not sensitive to the geometric location of the
dusty particles. For example a spherical envelope or a disky geometry
with identical density distributions will - with some dependence on
the optical depth - both have the same temperature distribution and
thus spectrum. It is thus hard to disentangle the geometry from the
observed SED, i.e.  the SED fitting is degenerate, and to break this
degeneracy we need additional information.

To conclude, the high spatial resolution observations provide the
geometry of the circumstellar material, but do not provide information
on its astrophysical properties. The SED fitting can provide that
information, but fails in determining the geometry. It may be clear
that combining the two is the way forward in characterizing the
material around MYSOs.

\begin{figure}
\begin{center}
  \includegraphics[width=0.6\textwidth]{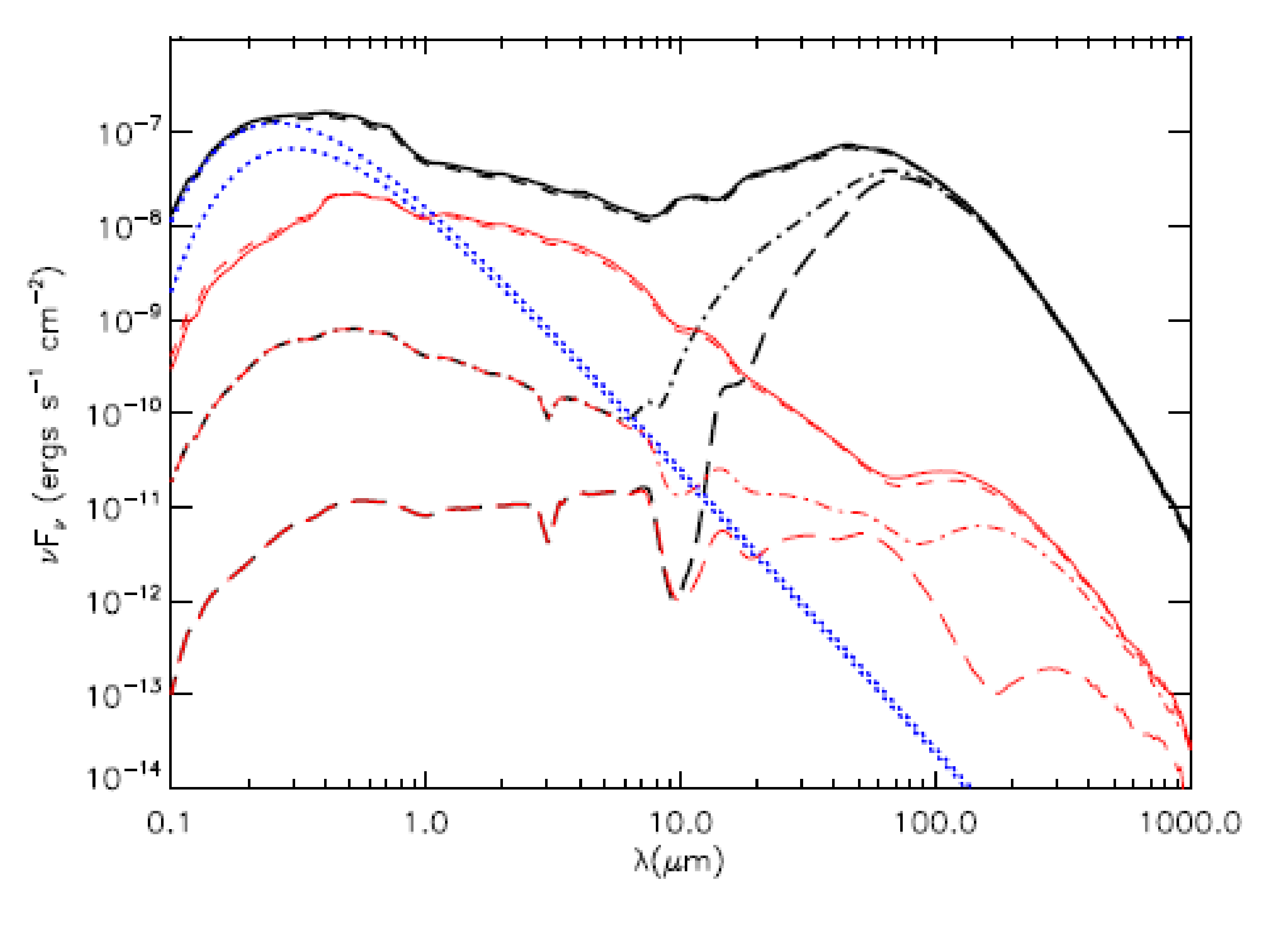}
  \caption{The Spectral Energy Distribution for an MYSO model seen at
    various inclination angles. The dotted lines denote the stellar
    input models, while the observable SED is shown for different
    inclination angles. The thicker lines represent the total flux,
    the thin red lines represent scattered light. The top model is the
    face-on situation where we look inside the cavity, while the
    bottom model shows the edge-on case. The face-on model is brighter
    in general and shows more near-infrared emission, probing the
    inner regions, which are obscured from view in more edge-on
    situations. The edge-on models display much stronger silicate
    absorptions at 10$\mu$m for the same reason.  Figure taken from
    Zhang \& Tan (2011).}
\end{center}
\end{figure}

\subsection{Two-dimensional models}

Above we outlined the need for spatially resolved observations to
break the degeneracy in SED modelling. This poses a number of
challenges, some of them obvious, and some of them are perhaps less
expected.

Firstly, the number of possible geometries that one can choose to
model is unlimited, indeed, it would be limited only by our
imagination. However, some geometries are more likely than others and
this ``free parameter'' will have to be chosen {\it a priori}. In the
case of Massive Young Stellar Objects, the paradigm of a spherically
symmetric envelope with cavities carved out by a bi-polar flow, and a
central star surrounded by a dense circumstellar disk is fairly well
established (see also Figure~1). A widely used dust code which
incorporates these elements is that by Whitney et al. (2003), while a
more recent example is that of Zhang \& Tan (2011). These codes allow
the user to specify many geometrical parameters, some of which can be
derived from other observations. The opening angle of the cavity can
be found from larger scale scattered light observations in the {\it K}
band (\cite{dewit_2010}), while the inclination of the system can be
inferred from other observations, such as for example CO rotational
emission line imaging at mm wavelengths. Figure~3 demonstrates some of
the dependencies involved. It shows the emergent SED for the same dust
model, which is seen at different inclinations. The resulting SEDs are
markedly different, most notably, much more absorption occurs in the
line of sight for the edge-on model, while the hot inner parts traced
by the near-infrared emission are revealed in the face-on model.

Secondly, at the moment it is expensive to compute for a given
geometry both the emergent SED and also the resulting image. It is
therefore not feasible to probe the entire (free-)parameter space in
the same manner as described above for the SED fitting, and choices
have to be made to make the fitting process computationally more
economical.

Thirdly, and perhaps surprisingly, it is not trivial to arrive at
formal best fits when combining the spectral energy distribution and
spatial information. Consider for example that a typical, well
sampled, SED has of order 10-15 photometric points covering the range
from the optical and near-infrared to millimetre wavelengths. A
simple, single baseline visibility spectrum covering 8-13 $\mu$m such
as shown in Fig.~2 already has more independent datapoints, and a
formal reduced $\chi^2$ would put less weight, if at all, to the SED,
despite the fact that it probes a much larger temperature range than
the MIDI spectrum.

We probably should point out at this stage that not many papers have
had the luxury to combine excellent SED information with high spatial
resolution data to constrain the circumstellar environments of
MYSOs. The practice thus far is well summarized and explained in de
Wit et al. (2010, 2011). In these papers, the authors first fix as
many input parameters as possible by using all available information
on the sources such as the stellar parameters and system
inclinations. They also determine the cavity opening angle by using
existing images in the near-infrared which probe the light scattered
off the inner cavity walls. They then first consider the best fitting
SED models from the grid computed by \cite{Robitaille}. This limits
the number of models that require the computation of model images. It
should be noted here however, that many models fitting the SED do not
match the spatial information. This is presumably due to the limited
parameter space probed, but this approach does help narrowing down the
possibilities.  These authors then put equal weight to the information
provided by the SED and the MIDI interferometric data respectively and
looked for various models that simultaneously fit the SED and
visibilities. A final check on the best fitting models was then done
by comparing the model output at longer, mm, wavelengths with
spatially resolved data. An example best fitting model is shown in
Fig.~4 for the object AFGL 2136. The surprising finding in this case
is the discovery of an unresolved central source that is strongly
emitting in the mid-IR wavelength regime. This is hypothesized to be
either a small, dense accretion disk or a bloated star whose extended
atmosphere is the result of strong accretion onto the star.

\begin{figure}
\begin{center}
  \includegraphics[width=0.6\textwidth]{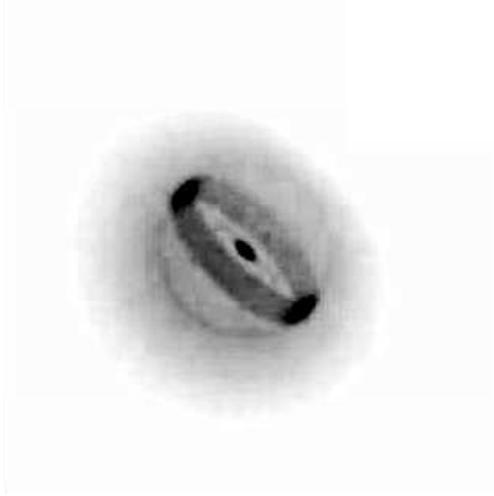}
  \caption{An image at 10$\mu$m of a model which both fits the SED and
    interferometry for the MYSO AFGL 2136. The source consists of a
    bright, unresolved, structure in the center, which is surrounded
    by a dusty torus that is oriented in the NW-SE direction. The
    bi-polar flow shows a blue-shifted component in the South-Eastern
    direction and a red-shifted component to the North-West. The
    red-shifted component will be more obscured, as can also be seen
    in this figure, where the northern component is not visible.
    Shown is AFGL 2136 as per de Wit et al. (2011).}
\end{center}
\end{figure}

\section{Final Remarks}

Remarkable progress has been made in star formation and the study of
Massive Young Stellar Objects over the past decade. This is in no
small part due to the interferometric facilities having become
mainstream and the observations themselves routine.  So far however,
the number of interferometrically observed objects is still limited,
while the number of objects for which model independent images are
available is even smaller. 

The future holds much promise, the continually improving sensitivities
will allow us to observe significantly more objects.  In addition,
instruments that can combine more baselines such as those on VLTI
(PIONIER, MATISSE, combining 4 beams), CHARA (up to 6 beams), and MROI (10
elements),  are either just installed on the telescope, or in an
advanced stage of development and construction. This will make it much
easier to obtain images and pave the way for statistical studies as it
will be possible to begin studying the MYSOs observed in this manner
as function of their properties such as evolutionary state and mass.

The improved sensitivites will also allow us to move away from
continuum work alone and enable studies of the emission lines. This
will provide valuable information on the stellar parameters, accretion
flow and bi-polar jets (e.g. Davies et al. 2010).  Last but not least,
the synergy with ALMA, which can achieve similar resolution, but
probes cooler material, allows us to ``map'' the circumstellar
environment of MYSOs from very close to the star to larger scales at
unprecedented detail.

\section*{Exercises}

\begin{enumerate}

\item

Infrared emission is due to dust heated by the star's radiation. The
further dust is from the star, the cooler it is. Relate the distance $d$
of a dust grain to a star as a function of the star’s radius $R_*$,
temperature $T_*$ and the dust grain’s temperature $T_d$. For this
derivation, you can assume that the star radiates like a black body,
and that a spherical dust particle absorbs 100\% of the energy that
falls on it, which it then re-radiates as a black body.

\item

A typical Massive Young Stellar object has $T_*$ = 35000K and $R_*$ =
8.4 solar radii, and is at a distance of 3.8 kpc. Now, take the dust
grain at its sublimation temperature, say 1000K, and work out the
angular distance of this dust grain to the star.

\item

Most such objects are too faint for AMBER. With the help of Wien's 
displacement law estimate which size scales and temperatures we can
probe with MIDI.

\item

We can now also check why T Tauri stars are rarely resolved with OIR
interferometry. Look up the stellar parameters and distance typical for a T
Tauri star and evaluate the size of the inner dust disk. 

\end{enumerate}




\end{document}